\documentclass[12pt,preprint]{aastex}

\def\plotfiddle#1#2#3#4#5#6#7{\centering \leavevmode
    \vbox to#2{\rule{0pt}{#2}}
    \includegraphics{#1}}

\newcommand{\Mstar}{M_*}

\newcommand{\Msun}{M_{\odot}}
\newcommand{\Rin}{R_{\rm in}}
\newcommand{\Rout}{R_{\rm out}}
\newcommand{\Rhole}{R_{\rm hole}}
\newcommand{\Rinner}{R_{\rm inner}}
\newcommand{\Rstar}{R_*}
\newcommand{\Rsun}{R_\odot}
\newcommand{\Msunperyr}{M_{\odot}\,{\rm yr}^{-1}}
\newcommand{\Lsun}{L_{\odot}}

\newcommand{\nH}{n_{\rm H}}

\newcommand{\percc}{\rm \,cm^{-3}}

\newcommand{\persqcm}{\rm \,cm^{-2}}
\newcommand{\gpersqcm}{\rm \,g\,cm^{-2}}

\newcommand{\kms}{\,{\rm km}\,{\rm s}^{-1}}

\def\micron{\hbox{$\mu$m}}

\shorttitle{CO Emission from V836 Tau}
\shortauthors{Najita et al.}

\begin{document}


\title{CO Fundamental Emission from V836 Tau\altaffilmark{1}}


\author{Joan R. Najita, Nathan Crockett}
\affil{National Optical Astronomy Observatory, 950 N. Cherry Ave., 
Tucson, AZ 85719} 

\and

\author{John S. Carr}
\affil{Naval Research Laboratory, Code 7213, Washington, DC 20375}

\altaffiltext{1}{The data
presented herein were obtained at the W.M. Keck Observatory from 
telescope time allocated to NASA through the agency's scientific 
partnership with the California Institute of Technology and the 
University of California.  The Observatory was made possible 
by the generous financial support of the W.M. Keck Foundation.}



\begin{abstract}
We present high resolution $4.7\micron$ CO fundamental spectroscopy 
of V836 Tau, a young star with properties that are between those of 
classical and weak T Tauri stars and which may be dissipating its 
circumstellar disk.  
We find that the CO line profiles 
of V836 Tau are unusual in that they are markedly double-peaked, even 
after correcting for stellar photospheric absorption in the spectrum.  
This suggests that the CO emission arises from a restricted range of 
disk radii ($< 0.5$\,AU), in contrast to the situation for 
most classical T Tauri stars where the CO emission extends 
out to much larger radii ($\sim 1-2$\,AU).  We discuss whether 
the outer radius of the emission in V836 Tau results from the 
physical truncation of the disk or an excitation effect.  We also 
explore how either of these hypotheses may bear on our understanding 
of disk dissipation in this system.  
\end{abstract}





\section{Introduction}

V836 Tau is an interesting young star with properties that are 
between those of classical and weak T Tauri stars. 
Its optical spectrum displays relatively weak, variable H$\alpha$ 
emission, with measured equivalent widths in the range 
1--25\AA\ (White \& Hillenbrand 2004; 
Beristain, Edwards, \& Kwan 2001; Kenyon et al.\ 1998; 
Wolk \& Walter 1996; Mundt et al.\ 1983).  
Similarly, V836 Tau also shows little veiling and has one of the 
smallest measured stellar accretion rates among T Tauri stars 
(Hartigan, Edwards, \& Ghandour 1995). 
Nevertheless, the detection of H$\alpha$ emission with an 
inverse P Cygni profile (Wolk \& Walter 1996) demonstrates the 
continued, possibly intermittent, accretion of material onto the star. 
Millimeter wavelength CO emission has also 
been detected from V836 Tau (Duvert et al.\ 2000)  
despite the relatively low disk mass $\sim 0.01\Msun$ 
inferred from the submillimeter dust continuum properties 
(e.g., Andrews \& Williams 2005).

At infrared wavelengths, V836 shows little or
no excess in the $K$- and $L$-bands, but a much stronger excess
indicative of an optically thick disk is apparent 
beyond 10$\micron$ (Strom et al. 1989; Skrutskie et al. 1990; 
Padgett et al.\ 2006).  
Interpreting the unusual spectral energy distribution (SED) 
and low stellar accretion rate of V836 Tau as indications that 
the system is on the verge of dissipating its disk, 
Strom et al.\ (1989) consequently referred to 
V836 Tau as a ``transitional T Tauri star'', a term reflecting 
the view that classical T Tauri stars
evolve into weak T Tauri stars with the dissipation of the disk
playing a central role in that evolutionary process. 
More generally, transition objects, 
systems that can be modeled as an optically thick 
disk that has an optically thin region (a hole or a gap) 
at smaller radii, 
have been suggested to be in the process of 
dissipating their disks from the inside-out (Strom et al.\ 1989; 
Skrutskie et al.\ 1990), possibly as a result of 
the formation of planetesimals (Strom et al.\ 1989), 
a giant planetary companion (e.g., Skrutskie et al.\ 1990), 
disk photoevaporation (Clarke et al.\ 2001; Alexander et al.\ 2006), 
or potentially other processes.

We report here on the properties of the 4.7$\micron$ CO fundamental 
($\Delta v$=1) emission from V836 Tau to see whether it can 
provide any insights into the extent and nature of disk dissipation 
in the system. 
CO fundamental emission has been previously characterized as a probe 
of circumstellar disk gas (Najita et al.\ 2003, 2007a; 
Blake \& Boogert 2004; Brittain et al.\ 2007).
In the case of disks surrounding low mass stars, 
CO fundamental emission is estimated to arise from 
within $1-2$\,AU of the star (Najita et al.\ 2003). As a result, 
the emission is well-suited to probing the radial distribution 
of gas in the terrestrial planet region of the disk. 
Recent thermal-chemical models of the inner regions of disks 
surrounding low mass stars suggest that the 
CO emission arises from the warm surface region of the disk that 
is heated by stellar (X-ray) irradiation and possibly mechanical 
heating processes (e.g., Glassgold et al.\ 2004). 
The CO fundamental emission from V836 Tau was discussed briefly in 
an earlier study (Najita et al.\ 2003).  
Here we report a new, higher signal-to-noise observation of the CO 
emission, which we use to characterize the radial distribution 
of the gas in inner $\sim$1--2\,AU of the disk surrounding V836 Tau.

\section{Observations}

We obtained a high-resolution ($R$=25,000) spectrum of V836 Tau on the
night of 2003 January 11 at the W.M. Keck Observatory using NIRSPEC 
(McLean et al.\ 1998) 
in echelle mode with a 0.43$^{\prime\prime}$ slit. 
Our spectral setup included two echelle orders covering 
the spectral ranges $4.586-4.662\mu$m (in order 16) and 
$4.890 - 4.972\mu$m (in order 15).  These wavelength regions include 
low-$J$ CO $v$=1--0 R-branch transitions in order 16 and high-$J$ CO
$v$=1--0 P-branch transitions in order 15.
Observations at $4.7\mu$m are dominated by strong 
thermal background continuum and telluric emission lines. In order to
remove these features, the telescope was nodded 
in an ABBA pattern at 2 minute intervals 
between two positions in the slit separated by 12$\arcsec$. 
During our observations, the seeing varied between
$\sim$0.6$^{\prime\prime}$ and $\sim$1.1$^{\prime\prime}$ mainly
as a result of high wind.  As a result, the total effective integration 
time on the source was 40 minutes.  In addition, fixed pattern noise on
the left two quadrants of the detector increased the 
noise in these quadrants 
by a factor of $\sim$1.5 relative to the noise on the right
side of the array.  This reduced the signal-to-noise ratio 
in the short-wavelength half of each order.

All data reduction procedures were carried out using IRAF. 
Subsequent images were subtracted from each other, removing the
background and telluric emission lines to first order. 
A flat field correction was applied to each differenced image.
To minimize the impact of imperfectly cancelled telluric emission
lines on the extracted spectrum, we rectified the images so
that the image of the slit ran vertically along pixel columns.  The
required geometric transformation was obtained from the original 
pattern of telluric
emission lines using the IRAF tasks FITCOORDS and TRANSFORM.
Individual differenced images were average combined and one-dimensional
spectra were extracted for each nod position and order.  
Because the spectra had been rectified, large background apertures 
(located on both sides of the source aperture) 
could be used in extracting the spectrum, producing a 
higher signal-to-noise spectrum.

Telluric absorption features were removed by dividing the program
spectrum by that of a telluric standard 
(the hot star HR 1605; spectral type A8), which  
was obtained and reduced in the same way as the V836 Tau spectrum.  
The spectrum of HR 1605 was mostly featureless except for absorption 
in the 7--5 Pf$\beta$ line.  Consequently, the divided spectrum has 
artificially enhanced hydrogen emission in this region. 
To correct for the difference in airmass between the object 
and the standard spectra, 
we used the IRAF task TELLURIC to perform the division. 

Telluric absorption lines were used for the wavelength calibration. 
An approximate flux calibration was obtained by acquiring additional
spectra of both V836 Tau and HR1605 with the 0.72$^{\prime\prime}$ 
NIRSPEC slit.
Given the poor seeing, slit losses are likely to have been 
significant. 
The short integrations used for the V836 Tau observations result 
in a formal 1-$\sigma$ error of 25\% in the flux calibration of 
each order.  The possibility of slit losses contributes 
additional uncertainty in the flux. 

Nevertheless, the resulting continuum fluxes for each order 
agree to $\sim 15$\% with the $4.5\micron$ Spitzer/IRAC flux reported 
for V836 Tau (Padgett et al.\ 2006; [$F_{4.5}=0.127$ Jy]).
For comparison, our calibrated continuum flux of 0.128 Jy, the 
average of the continuum level in orders 15 and 16, is 
$\sim 80$\% brighter than the value 
of $\sim 0.07$ Jy that we reported previously for V836 Tau in the 
same wavelength region based on NIRSPEC data taken in 2001 
(Najita et al.\ 2003).
It is also 30\% fainter than the 0.18 Jy continuum flux at $4.78\micron$ 
that is implied by extrapolating the Spitzer/IRS spectrum of V836 Tau 
(Furlan et al.\ 2006)
to that wavelength.
Because of the relatively large uncertainty in the flux calibration 
for each order, we are unable to determine if there is a shallow 
continuum slope between the two orders. 
We therefore adopted the continuum flux slope measured by Furlan et al.\ 
and scaled the flux so that the average of the continuum fluxes at the 
wavelength centers of orders 15 and 16 agrees with our measured 
average value. 

We also acquired on 2004 November 23 similar NIRSPEC spectra of the 
K5V star 61 Cyg A.  These spectra were used in calibrating 
synthetic models of the stellar photospheric 
contribution to the V836 Tau spectrum (see section 3.1).  
The data were reduced and wavelength calibrated with the same 
procedure used for the V836 Tau data.

\section{Results}

The resulting CO fundamental spectrum is of significantly higher
signal-to-noise than our previously reported spectrum of V836 Tau 
(Najita et al.\ 2003).  In the earlier spectrum, CO emission was
clearly detected, but the shape of the line profile was unclear and 
the properties of the emission were unusual. 
The line widths of the low-$J$ R lines were found to be significantly 
narrower than the widths of the high-$J$ P lines 
($41\kms$ compared to $58\kms$). 
In addition, 
the emission at $4.6\micron$ appeared to peak at a radial velocity
redward of the stellar velocity.  A comparison with the 
new spectrum shows that this was because the blue component of the 
low-$J$ profiles was not recovered in the earlier data, a result 
of our limited ability to correct for telluric absorption given 
the lower signal-to-noise of the earlier data.  
The new, higher signal-to-noise data show $v$=1--0 emission that is 
double-peaked in both the $4.6\micron$ and $4.9\micron$ regions 
(Fig.~1 and 2, respectively, top panels) and centered at the 
radial velocity of the star.  The emission equivalent width is 
comparable to that in the earlier spectrum. 

The double-peaked line profiles of the CO $v$=1--0 emission 
from V836 Tau are unusual in that the majority of classical 
T Tauri stars that have been studied to date show CO $v$=1--0 
line profiles that are centrally peaked (Najita et al.\ 2003). 
For emission 
arising in a disk, double-peaked profiles indicate that the emission 
arises from a limited range of disk radii.  For example, this is the 
interpretation given to the double-peaked 
$2.3\micron$ CO overtone ($\Delta v$=2) emission lines 
that are observed from actively accreting young stars (e.g., WL 16, DG Tau;
Carr et al.\ 1993; Najita et al.\ 1996, 2000).

\subsection{Stellar Photosphere Component}
Because the observed spectrum is a composite of the emission from
the star and disk, some fraction of the central dip in the CO line
profile may result from absorption in the stellar photospheric
component.  
We estimated the stellar contribution to the spectrum using the recent
version of the stellar spectral synthesis program MOOG (Sneden 1973). 
As inputs, we used the NextGen model atmospheres (Hauschildt et al.\
1999) and the CO linelist of Goorvitch (1994). The ability to fit
the CO spectrum in this wavelength region was verified on a spectrum of
the K5 dwarf 61 Cyg A, which was obtained with the same instrumental
setup as the spectrum of V836 Tau. For our adopted stellar parameters
for 61 Cyg A
($T=4400$\,K, $\log g = 4.5$) and near-solar metallicities, we obtain an
excellent fit to the spectrum. In contrast, with the atmospheric models
of Kurucz (1993), it is not possible to match the relative strengths of
the weak and strong CO lines.

We selected an atmosphere model that is appropriate for the stellar 
photospheric temperature ($T_{\rm eff}= 4000$\,K) and gravity 
($\log g = 4.0$) that are implied by the stellar effective temperature
and luminosity of V836 Tau measured by White \& Hillenbrand (2004).  
We adopted a solar metallicity, as appropriate for the Taurus star
forming region (Padgett 1996; Santos et al.\ 2008).

A comparison of the synthetic stellar spectrum and the observed 
spectrum of V836 Tau shows that much of the structure in the 
continuum is due to absorption features in the stellar photosphere  
(Figs.\ 1 and 2, top panels).  
Stellar photospheric features are expected to be detectable in 
high signal-to-noise spectra given the weak near-infrared excess 
of V836 Tau. 
The heliocentric radial velocity of $18.5 \kms$ 
(which corresponds to a topocentric velocity of $v_{\rm obs}=35.5 \kms$) 
and the stellar rotational velocity of 
$v\sin i \simeq 12.1 \kms$ measured by 
White \& Hillenbrand (2004)
are in agreement with the 
stellar photospheric features identified in the observed spectrum. 
These parameters, combined with a continuum veiling of $r_{4.7}=2$ 
(where $r_{4.7}$ is the ratio of the excess flux to the stellar 
photospheric flux at $4.7\micron$),  
provides a reasonably good fit to the observed spectrum.  
The depth of the CO absorption in the veiled stellar photospheric 
spectrum shows that some of the central absorption in the observed 
CO emission profile arises from CO absorption in the stellar 
photosphere.

The measured veiling has two sources of uncertainty.  The 
low signal-to-noise of the spectrum introduces uncertainty in the 
location of the continuum, contributing an uncertainty of 
$\pm 0.3$ to the veiling measurement.  The uncertainty in the 
stellar effective temperature ($\pm 200$\,K) and $\log g$ 
($\pm 0.3$) of V836 Tau (see \S 3.3) contributes an additional 
uncertainty of $\pm 0.16$ to the veiling. 

Figure~3 shows the dereddened SED of V836 Tau.  We constructed the SED from the 
$UBVRI$ photometry of Kenyon \& Hartmann (1995), 
the $JHK$ fluxes from 2MASS, and 
the {\it Spitzer} IRAC and MIPS photometry reported by Padgett et al.\ (2006). 
The fluxes were dereddened using the reddening law of Mathis (1990) 
assuming $A_V=1.1$ as in Furlan et al.\ (2006).
A Basel stellar atmosphere v2.2 (corrected; Lejeune 2002) 
with $T_{\rm eff} = 4060$\,K and $\log g = 4.0$ 
provides a good fit to the dereddened optical colors of the star. 
The stellar contribution as shown is roughly consistent with 
the minimal veiling at $R$ and $I$ measured by White \& Hillenbrand (2004).
Overall, the SED is approximately photospheric below $3.5\micron,$
with a significant infrared excess beyond that wavelength. 

Our measured veiling of $r_{4.7}=2$ is consistent with values that are
implied by the stellar photospheric contribution to the SED shown in 
Figure 3 and photometric measurements of V836 Tau in the literature. 
For consistency with the stellar photospheric contribution to the 
SED, our veiling requires a continuum flux of $\sim 0.16$ Jy
at the time our spectrum was obtained, which is within the level of 
uncertainty in our flux calibration.

\subsection{CO Emission Line Profile}

To obtain the disk contribution to the CO fundamental spectrum, 
we subtracted 
the veiled stellar photosphere from the observed spectrum (Figs.\
1 and 2, bottom panels).  The difference spectrum shows that even
after accounting for stellar photospheric absorption, the 
residual CO emission spectrum has a double-peaked line profile. 
We can better define the line profile by averaging multiple 
CO emission lines to improve the signal-to-noise of the line 
profile.  
In constructing an average line profile, regions of 
poor telluric correction 
(below a cutoff value of 90--95\% transmission) 
were excluded from the average. 
The regions to exclude were determined from the observations of 
the telluric standard.

The average profiles of the low-$J$ R and high-$J$ P lines 
are shown in Figures 4 and 5 (solid black lines),
respectively.
In these figures, the individual lines that contribute to the 
average are shown as histograms.  The points 
that are included in the average are indicated by asterisks. 
The distribution of the asterisks shows that many of the 
deviant points are excluded by the weighting scheme, as 
desired. 
For both the low-$J$ R and high-$J$ P lines, the average line profile 
indicates that the CO emission extends to approximately $\pm 90\kms$.  
The location of the ``horns'' of the profile (at approximately 
$\pm 30-45\kms$) further indicates
that the CO emission arises 
from a limited range of disk radii ($\Rout/\Rin \sim 5-9$).

\subsection{Disk Emission Model}

To estimate the range of disk radii from which the emission originates  
and the excitation conditions therein, we modeled the 
residual (stellar photosphere-subtracted) 
CO emission using a simple model of emission from 
a gaseous disk under the assumption of Keplerian rotation
and thermal 
level populations (e.g., Carr et al.\ 1993; Najita et al.\ 1996). 
Given the estimated extinction of 
$A_V=1.1$ (Furlan et al.\ 2006)  
we assumed a negligible extinction at $4.7\micron$. 
For simplicity, no underlying continuum was assumed in fitting 
the line emission. 
This is reasonable given our limited goal 
of estimating the radial extent of the emitting gas. 

An important model parameter is the stellar mass. 
We can estimate the stellar mass of V836 Tau both 
from pre-main-sequence 
evolutionary tracks and from available dynamical mass estimates 
of pre-main-sequence stars of the same spectral type and similar age.
The K7 $\pm 1$ spectral type (or T$_{\rm eff} = 4000\pm 200$\,K; 
White \& Hillenbrand 2004) 
corresponds to a mass of $0.71^{+0.24}_{-0.18} \Msun$ using the 
evolutionary tracks of Siess et al.\ (2000). 
Similarly, other T Tauri stars in Taurus with K7 spectral types 
have measured dynamical masses (Simon et al.\ 2000) of 
$0.72\Msun$ (DL Tau) and $0.84\Msun$ (GM Aur).
Thus, the K7 dynamical masses 
and HR diagram position of V836 Tau are consistent with a 
stellar mass of $0.7-0.8 \Msun$.  If the spectral type 
is larger or smaller by one subclass, a larger 
range in mass is allowed ($0.5-1.0\Msun$).  Here, we assume 
a stellar mass of $0.75\Msun$.

An additional model parameter is the system inclination, which 
we can obtain from the stellar rotation period $P_{\rm rot}$
of $6.76$ days determined from long-term monitoring 
(Grankin et al.\ 2008), the projected 
stellar rotational velocity 
$v_*  \sin i$ ($12.1 \kms$; White \& Hillenbrand 2004), and 
the stellar radius $\Rstar.$  These are related by 
$$ v_* \sin i = {2\pi \Rstar \over P_{\rm rot}} \sin i.$$
The stellar radius can be estimated from the measured 
stellar luminosity and temperature. 
The stellar luminosity, obtained by integrating the stellar 
component in the SED fit (Fig.~3; see also Furlan et al.\ 2006), 
is $0.58 \Lsun$ at the average Taurus distance of 140\,pc, 
with a range of $0.46 - 0.87\Lsun$ if we account for the 
range in distances ($126-173$\,pc) derived to individual Taurus 
objects (Bertout \& Genova 2006). 
For $T_{\rm eff}=4000 \pm 200$\,K,  
$\Rstar= 1.6^{+0.40}_{-0.22}\Rsun,$ where the uncertainty in the 
stellar temperature dominates the error on the low luminosity end, 
and the uncertainty in the distance dominates the error on 
the high luminosity end.  
An examination of all of the uncertainties in the above properties 
gives a possible inclination range of $55 - 90$ degrees. 
Since V836 Tau does not have the extreme colors of an edge-on disk 
system (e.g, D'Alessio et al.\ 2006), we assume a more modest 
inclination of 65 degrees for the modeling. 

With these constraints on the stellar mass and system inclination, 
we can then infer the range of disk radii over which the emission 
arises assuming Keplerian rotation. 
The maximum velocity extent of the $v$=1--0 CO emission lines 
($v_{\rm max} \sim 90\,\pm 10\kms$; see sec.~3.2)  
and the inclination range of $i=55-90$ degrees 
implies an inner radius of $\Rin = 0.05-0.09$\, AU 
for the emitting gas. 
This range of $\Rin$ is consistent with the values 
of $\Rin$ measured for classical T Tauri stars using 
CO emission line profiles (Najita et al.\ 2007a; Carr 2007).
The overall shape of the emission line profile further suggests that 
the emission extends out to a radius 
$\sim 6 \Rin$, significantly 
less than the $>20 \Rin $ that would be 
inferred from the centrally peaked CO fundamental emission 
profiles of most T Tauri stars (Najita et al.\ 2003). 
Thus, the outer radius of the V836 Tau CO emission is in the range 
$0.3-0.5$\,AU.

We can provide constraints on the mean conditions in the emitting gas
by modeling the CO spectrum with a radially constant
excitation temperature and column density.
The similar shape and strength of the $v$=1--0 
emission lines over a wide range in $J$ reveals that the emission 
lines are optically thick, i.e., that the gas column density is 
$> 0.001\gpersqcm$ if turbulent line broadening is negligible.  
In addition, the absence of detectable $v$=1--0 $^{13}$CO emission lines 
limits the column density to $< 0.03\gpersqcm$ 
for a CO abundance of $3\times 10^{-4}$ relative to hydrogen and 
an interstellar $^{13}$CO/$^{12}$CO ratio of 90.

The weakness of the $v$=2--1 and $v$=3--2 lines in the spectrum 
requires either a low average excitation temperature or that the vibrational 
levels above $v$=1 depart significantly from thermal equilibrium.  
For LTE level populations, 
the relative strengths of the $v$=1--0 lines and the limit on the
strength of the $v$=2--1 transitions constrain 
the mean excitation temperature to the range 700--1100\,K.
Smaller excitation temperatures 
require larger emitting areas to produce the required line flux, as 
well as larger inclinations in order to produce the emission over 
the required range of velocities.  As a result, at temperatures 
$\lesssim 700$\,K, the requirement on the line flux drives the 
emitting radii to values large enough that the required velocities 
cannot be obtained at any inclination.  For a radially constant 
excitation temperature 
$\gtrsim 1200$\,K, the $v$=2-1 lines are too strong 
relative to the $v$=1--0 lines. 
In addition, given the constraints on the radial range (and therefore 
the emitting area) of the emission, the strength of
the $v$=1--0 lines are overpredicted at such high temperatures.

We can obtain a better fit to the average line profiles by allowing the
temperature to vary as a function of radius.
Consistent with the above considerations, 
a gas temperature that varies slowly with 
radius ($T = 1200\,{\rm K} (r/\Rin)^{-0.30}$) and 
a line-of-sight disk column density of 
$\Sigma = 0.003\gpersqcm$ 
that is radially constant between 
$\Rin= 13 \Rsun$ and $\Rout = 5 \Rin$, 
gives a reasonable fit to 
the spectra assuming LTE level populations (Fig.~6).
In addition to providing a reasonable fit to 
the relative line strengths of the $v$=1--0 lines, 
the model also reasonably fits the average line profiles of the 
$v$=1--0 lines (Figs.~4 and 5; heavy dashed line).

We can also fit the spectra with a model that uses a steep 
temperature gradient rather than a specified outer radius 
to limit the radial extent of the emission. 
With a temperature profile 
$T=1400\,{\rm K}(r/\Rin)^{-0.6}$  
and a radially-constant line-of-sight column density 
$\Sigma=0.0037 \gpersqcm$ and $\Rin=16.2\Rsun$, 
the 1--0 emission decreases sharply beyond $7-8\Rin$ 
because the Planck function contributes little at $4.7\micron$ 
at the low temperatures achieved at these radii 
($\lesssim 400$\,K).  
The high-$J$ P-branch lines are also optically thin at 
these radii.
The model provides a reasonable fit to the relative strengths 
of the $v$=1--0 lines.  The average line profiles are reasonably 
well fit (Figs.\ 4 and 5; heavy dash-dot line), 
although the central dip in the $v$=1--0 R lines 
is shallower than observed.  This is because the low-$J$ R lines 
remain optically thick (and continue to produce emission) to larger 
radii than the high-$J$ P lines.  In contrast, the outer 
radius to the emission used in the first model, produces  
similar line profiles for the low-$J$ R and high-$J$ P lines. 

To summarize, the observed spectra can be explained 
with an abrupt truncation of the CO emission 
beyond an outer radius of $7-8\Rin$. 
A steep temperature gradient where the temperature drops 
to $\lesssim 400$\,K at $7-8 \Rin$ is an alternative explanation.  
Note that there is a slight inconsistency in the model 
parameters used in the above fits.  The adopted inclination 
formally implies a distance of 165\,pc rather than the 
nominal Taurus distance of 140\,pc that we have assumed 
in the fits.  If V836 Tau is located at a larger distance 
than 140\,pc, the CO emission can be fit with similar model 
parameters to those used above, with the modification that 
$\Mstar = 0.88\Msun$ and the emission arises from radii 
1.18 times larger than assumed above.  
The higher mass is within the mass range allowed by the 
uncertainty in the stellar spectral type.

\section{Discussion}

\subsection{Truncated Excitation or Truncated Disk?} 

The CO emission from V836 Tau shows similarly double-peaked 
line profiles for lines spanning a large range in excitation 
temperature.  The line profile shapes indicate that the 
CO emission is truncated beyond $\sim 0.4$\,AU.  
The similarity in the relative strengths of the 1--0 transitions 
show that the 1--0 transitions are optically thick over the 
emitting region. 
As described in the previous section, 
these properties could be produced by a {\it physical} truncation 
of the gaseous disk beyond a radius of $\sim 0.4$\,AU or 
a truncation of the CO emission (but not the disk gas 
column density) at the same radius. 

Although it is difficult to determine which of these 
interpretations is correct based on the available information, 
some possibilities can be ruled out.  For example, 
the truncation of the CO emission is unlikely to result from 
the thermal dissociation of CO since the excitation 
temperature of the gas is much less than the thermal 
dissocation temperature of CO ($\sim 4000$\,K at the 
densities of inner disks). 

We might also consider the possible explanations that 
have been put forward for the origin of the 
double-peaked line profiles 
observed in the $v$=2--0 CO overtone bandhead emission 
from T Tauri stars and Herbig Ae stars at $2.3\micron$
(e.g., Carr et al.\ 1993; Najita et al.\ 1996, 2000). 
Such emission is detected only from sources with 
high accretion rates, probably a consequence of the  
temperatures and high column densities that are needed to produce the 
overtone emission.
Since these systems have disks that are believed to be 
radially continuous, the double-peaked lines are unlikely 
to arise from a radially truncated disk.   
We have previously described two possible explanations for 
the double-peaked lines that make up the CO $v$=2--0 bandhead:  
(1) the outer radius is a dust sublimation front, or  
(2) the transition from atomic H to H$_2$ at decreasing temperature 
depopulates the higher CO vibrational levels. 

In the first scenario, a dust sublimation front renders 
the line emitting layer optically thick in the continuum 
at temperatures below $\sim 1500$\,K, 
eliminating the contrast of the CO $v$=2--0 emission above the 
continuum at these radii (e.g., Carr 1989). 
This explanation is difficult to apply to the CO 
fundamental lines, because 
typical dust temperatures at the inferred 
outer radius of $\sim 0.4$\,AU for the CO fundamental emission 
are much below the dust sublimation temperature 
(e.g., D'Alessio et al.\ 1999) and measured dust sublimation 
radii are much smaller 
than the inferred outer radius of the CO fundamental emission 
(Eisner et al.\ 2005; Muzerolle et al.\ 2003).  
Therefore, the outer 
radius observed for the CO fundamental emission is 
unlikely to result from dust sublimation.  

As an alternative scenario, 
we previously speculated that 
the transition from atomic to molecular hydrogen at decreasing 
disk temperature leads to the depopulation of the higher 
vibrational levels of CO due to the lower collisional 
cross-section of molecular hydrogen with CO compared to 
that of atomic hydrogen with CO (Najita et al.\ 1996).
In chemical equilibrium, disks would 
transition from a mixture of CO and atomic hydrogen at small 
disk radii ($T > 2000$\,K) to a mixture of CO and molecular 
hydrogen at large disk radii ($T < 2000$\,K). 
In this situation, we estimated that radiative trapping would 
be able to maintain the $v \ge 2$ vibrational populations down 
to a temperature of $\sim 1500$\,K, below which significant 
depopulation would occur.
Using a non-LTE model of this kind, we were able to reproduce 
the line intensities and shapes of CO overtone emission lines 
covering a wide range of excitation conditions ($v$=2--0 to $v$=5--3)  
in the spectra of two Herbig Ae stars.  

The same effect is unlikely to apply in detail to the 
truncation of the CO fundamental emission seen in V836 Tau  
($\Rout \sim 0.4$\,AU). 
While chemical equilibrium may be relevant at the large 
column densities needed to produce the overtone emission,  
disks are expected to have significant vertical structure 
and to depart significantly from chemical equilibrium 
at the smaller column densities needed to produce the 
fundamental emission. 
Thermal-chemical models of T Tauri disks irradiated by stellar 
X-rays (Glassgold et al.\ 2004; Meijerink et al.\ 2008) 
indicate that overlying the large column densities 
where the overtone lines form is a warm surface layer 
that is expected to be 
conducive to the production of CO fundamental emission 
over a large range in radii (Glassgold et al.\ 2004).  

The X-ray irradiated disk models, which have currently studied 
the structure of disks over the region 0.25--2\,AU, find that throughout 
this range of radii disks possess a warm ($\sim 1000$\,K) 
surface layer ($\gtrsim 10^{21}\persqcm$) of CO mixed with 
atomic hydrogen (Glassgold et al.\ 2004; Meijerink et al.\ 2008).  
Thus, the CO fundamental transitions could plausibly 
be excited over radii within and beyond 1\,AU. 
This expectation is in agreement with observations of 
CO fundamental emission from T Tauri stars. 
Empirically, we find that 
almost all accreting T Tauri stars show CO fundamental emission 
and that the majority of CO fundamental emission profiles 
are centrally peaked.  This indicates that the emission 
arises from a wide range of disk radii 
($\Rout/\Rin > 20$, $\Rout \simeq 1-2$\,AU), 
without a sharp truncation in excitation 
as a function of velocity (Najita et al.\ 2003).

However, it is possible that these general trends, obtained 
for typical T Tauri stars, do not apply to V836 Tau.  
Whereas typical T Tauri stars have strong near-infrared 
excesses and stellar accretion rates $\sim 10^{-8}\Msunperyr$, 
V836 Tau has a weak near-infrared excess and a low 
stellar accretion rate $\sim 10^{-9}\Msunperyr$ 
(Hartigan et al.\ 1995 scaled to Gullbring et al.\ 1998; 
Herczeg et al.\ 2006). 
The weak near-infrared excess might indicate either 
a significant settling of grains out of the disk atmosphere 
or a lack of dust at small disk radii.  

In this situation, one might imagine that a lack of small 
grains could reduce the heating of the gaseous atmosphere,
if gas-grain collisions dominate the heating of the gaseous 
disk.  In a cooler atmosphere, vibrational CO emission would 
be more difficult to produce, perhaps contributing, thereby, 
to the truncation of the CO emission.  
This does not seem likely in the context of recent X-ray 
irradiated disk atmosphere models, in which the 
gas is heated directly by X-rays or accretion-related processes 
and the grains function primarily as a coolant for the gas 
through collisions.  
In such a model, we might expect grain growth and settling 
to produce warmer gaseous atmospheres, rather than a radial 
truncation of the CO emission. 

Alternatively, one might imagine that CO might be less abundant 
in a grain-poor disk atmosphere if the CO is synthesized from 
H$_2$ that forms on grains.  However, in the X-ray irradiated 
disk atmosphere models, the grains are typically warmer than 
100\,K within 1\,AU, and have therefore been assumed to play 
no significant role in the synthesis of CO and other molecules 
(Glassgold et al.\ 2004). 
As a result, grain growth and settling is not expected to 
significantly reduce the strength of the CO emission from the 
disk atmosphere. 

Another possibility is that the low accretion rate of V836 Tau 
compared to the average T Tauri accretion rate 
($\sim 10^{-8}\Msunperyr$; Hartmann et al.\ 1998)
might play a role in truncating the CO emission.  
Since accretion-related processes may heat the gaseous 
atmosphere (Glassgold et al.\ 2004), 
the gaseous atmosphere may experience reduced heating   
at the lower accretion rate of V836 Tau. 
While the available heating is clearly able to produce 
detectable $v$=1--0 CO fundamental emission from V836 Tau, 
could reduced heating radially truncate the emission, perhaps 
via a non-LTE effect? 

A simple estimate indicates that the vibrational populations
could be in non-LTE.
The critical density $n_{\rm cr}$ for the $v$=1 level of CO, obtained
from the Einstein A-values for the $v$=1--0 transitions (e.g.,
Goorvitch \& Chackerian 1994) 
and the $v$=1--0 collision rates for CO with atomic
hydrogen (Glass \& Kironde 1983), varies as $T^{-1/2}$ and equals
$\sim 5 \times 10^{12}\percc$ at a temperature of 1000\,K
(see discussion in Najita et al.\ 1996).
Our modeling of the CO emission from V836 Tau shows 
that the $v$=1--0 CO emission arises from a gas column density 
$\sim 0.01 \gpersqcm$ at disk radii $\lesssim 0.4$ AU.  In the 
D'Alessio et al.\ (1999) disk model, the density in the disk 
atmosphere over this range of column density is 
$\nH \sim 10^{11}-10^{12} \percc$ at radii 0.1--0.4 AU.  
Therefore, a rough estimate is $\nH/n_{\rm cr} \sim 0.1$ 
in the CO emitting region. 

Our modeling of the CO emission further shows that the 
$v$=1--0 CO lines are optically thick, with $\tau \sim 10$. 
Since the escape probability for Gaussian lines depends 
asymptotically on the line optical depth as $\sim \tau^{-1}$ 
(Mihalas 1978), the line optical depth approximately compensates 
for the low value of $\nH/n_{\rm cr}$ so that 
the $v=1$ level could be in LTE. 
The situation for the higher vibrational levels is possibly 
similar, with the critical densities for these levels being 
comparable to the critical density for $v$=1.
This because the A-values of the higher vibrational levels are
larger (proportional to $v$), but the collision rates may also
be larger because of the contribution from $\Delta v > 1$ collisions 
(A. Glassgold, personal communication).
Thus, the CO vibrational level populations could plausibly 
be in LTE, consistent with the assumption made in \S 3.3.  

However, both lower temperatures and lower densities in the CO 
emitting region will tend to drive departures from LTE.  Reduced 
accretion heating can produce both lower temperatures and lower 
densities.  As the accretion heating is reduced, the vertical temperature
inversion that puts the CO into emission will become increasing
limited to the lower column density surface region that can be heated
by external irradiation (e.g., by stellar X-rays).  This surface
region will be characterized by lower densities.
Detailed calculations of the 
thermal, chemical, and density structure of disk atmospheres 
are needed to address this issue quantitatively.

Because such calculations are currently lacking, we might 
take instead a more empirical approach and compare the 
CO fundamental line profiles of V836 Tau with 
those of other T Tauri stars with low stellar 
accretion rates and/or low near-infrared excesses.  
As an example of such a comparison, Figure 7 (top panel) 
shows the CO fundamental emission from LkCa\,15,  
a T Tauri star with a stellar accretion rate 
similar to that of V836 Tau 
(Hartmann et al.\ 1998; see also Najita, Strom, \& 
Muzerolle 2007b).
The spectrum shows the data reported in 
Najita et al.\ (2003), but re-reduced using the approach 
described in \S 2. 
The $v$=1--0 CO emission from LkCa\,15 is centered at the 
stellar velocity (short vertical lines, bottom panel), 
but compared to V836 Tau, 
it shows a centrally peaked profile. 

Although the signal-to-noise ratio of the spectrum is 
limited, weak stellar photospheric features appear to 
be present. 
In the bottom panel of Figure 7, we show the CO emission 
from LkCa\,15 after correction for a stellar photospheric 
contribution, following the approach described in \S 3.1.  
The stellar photospheric model assumes 
an Allard stellar atmosphere model with 
a gravity of $\log g = 4.0$, 
an effective temperature of 4400\,K to match the 
K5 spectral type of LkCa\,15 (Herbig \& Bell 1988), 
a stellar rotational velocity $v \sin i =12.5\kms$ 
(Hartmann, Soderblom, \& Stauffer 1987), 
and an observed (topocentric) radial velocity 
of $-39.8 \kms$, which is appropriate for  
the measured stellar radial velocity 
(Hartmann et al.\ 1987) and the observation date. 
A veiling of 2.5 times the stellar continuum roughly 
reproduces the strength of the stellar photospheric 
features near the 1--0 P30 line. 
This level of veiling is also consistent with the veiling at 
$5\micron$ implied by the SED (e.g., Furlan 
et al. 2006). 
Subtracting the veiled stellar photospheric component 
produces a disk CO emission profile that is even more 
centrally peaked (Fig.\ 7, bottom panel). 

In addition to its low stellar accretion rate, 
LkCa\,15 is also similar to V836 Tau in that it has the 
characteristics of a transition object 
with a weak near-infrared continuum and an 
optically thick outer disk (e.g., Bergin et al.\ 2004; 
Espaillat et al.\ 2007). 
Its properties therefore probe empirically how both the reduction 
in small grains and reduced accretion-related heating 
might affect the CO fundamental emission from the disk. 
The LkCa\,15 spectrum shows that in at least some 
cases, these effects do not radially truncate the CO 
fundamental emission within 1 AU.

To summarize, the double-peaked line profile of the CO fundamental 
emission from V836 Tau may indicate that the gaseous disk extends 
from close to the star ($\sim 0.05$\,AU) out to a physical 
truncation radius ($\sim 0.4$\,AU). 
If the gaseous disk in V836 Tau is instead continuous beyond 
this radius, the double-peaked CO 
profile would indicate a sudden truncation of the 
CO emission beyond a radius of 0.4\,AU. 
As discussed in \S 4.2, an abrupt decrement in excitation 
is not expected empirically, since  
the majority of CO emission profiles observed to date 
(both typical classical T Tauri stars and transition objects 
like LkCa\,15)
show centrally peaked CO profiles with no comparable 
decrement in excitation with radius.

However, a truncated emission profile may result from either 
an (anomalously) steep temperature gradient or departures from 
LTE that become important in low accretion rate systems.
These two possible explanations can be explored and 
potentially distinguished with a higher signal-to-noise spectrum 
that measures the strengths of the $v$=2--1 lines. 
A theoretical study of the possibility of non-LTE CO level populations 
would also be welcome in sorting out whether an excitation effect 
is a possible explanation for the outer radius of the emission.
Additional observations of low accretion rate sources 
would be useful in exploring this issue empirically. 

Most definitive of all would be observations of spectral line 
transitions that robustly probe the disk atmosphere of V836 Tau 
at excitation 
temperatures $< 400$\,K.  These observations would complement 
the insensitivity of the CO fundamental transitions to low 
temperature gas.  
If little emission is detected with these diagnostics at disk 
radii $>0.4$\,AU, that would strongly suggest that the gaseous 
inner disk in V836 Tau is physically truncated at $\sim 0.4$\,AU.
Some of the mid-infrared molecular emission diagnostics recently 
reported in the spectrum of AA Tau (C$_2$H$_2$, HCN, H$_2$O, OH; 
Carr \& Najita 2008) may prove useful in this regard.

\subsection{Nature of V836 Tau}

The possibility that the disk of V836 Tau is physically 
truncated beyond 0.4\,AU may bear on our 
understanding of the nature of the system.  
As noted in \S 1, V836 Tau has been previously classified as  
a transition object (Strom et al.\ 1989), 
a system with an optically thin inner disk (within $\Rhole$) 
and an optically thick outer disk (beyond $\Rhole$).  
The SEDs of these systems have been variously explained as 
a consequence of grain growth and planetesimal formation 
(e.g., Strom et al.\ 1989; Dullemond \& Dominik 2005), 
giant planet formation 
(e.g., Skrutskie et al.\ 1990; Marsh \& Mahoney 1992), 
or photoevaporation 
(e.g., Clarke et al.\ 2001; Alexander et al.\ 2006).   

While all of these processes can produce optically thin regions 
in the disk (inner holes or gaps), they make different 
predictions for stellar accretion rates and disk masses, 
as well as the radial distribution of the gaseous disk.  
As a result, measuring the radial distribution of disk gas 
and comparing the stellar accretion rates and disk masses 
of transition objects with those of accreting T Tauri stars of 
comparable age can potentially sort among the possible 
explanations for a transitional SED.  For example, 
a recent study using the latter approach showed  
that transition objects in Taurus (including V836 Tau) 
have stellar accretion rates that are on average $\sim 10$ 
times lower than those of non-transitional 
T Tauri stars with comparable disk masses (Najita et al.\ 2007b).  
Such a reduced stellar accretion rate 
is predicted for disks that have formed Jovian mass planets
(e.g., Lubow \& D'Angelo 2006), 
suggesting that giant planet formation may play a role in 
explaining the origin of at least some transition objects. 

Studies of the radial distribution of the gaseous disk in the 
system provide an additional way to distinguish the nature 
of individual transition objects. 
For example, although grain growth can render the inner disk 
optically thin (within $\Rhole$), the gas in the same region of the 
disk is not expected  to be altered significantly; gas would 
therefore fill the region within $\Rhole$. 
If a giant planet has formed with a mass sufficient to open a gap, 
both the gas and dust would be expected to be cleared dynamically 
from the vicinity of the orbit of the planet, 
creating an inner disk 
(within $\Rinner < \Rhole$) that is fed by accretion streams from an 
outer disk (beyond $\Rhole$; Lubow, Seibert, \& Artymowicz 1999; 
Bryden et al.\ 1999; Kley 1999; D'Angelo et al.\ 2003; 
Lubow \& D'Angelo 2006).  
Accretion streams are not expected in the case of a massive 
giant planet ($\sim 10 M_J$; e.g., Lubow et al.\ 1999), 
and no significant gas or dust is expected anywhere within 
$\Rhole$ in this case. 
A lack of gas or dust within $\Rhole$ is also expected 
in the photoevaporation case; such systems are further expected 
to have a very low disk mass 
($\sim 0.001 \Msun$; e.g., Alexander \& Armitage 2007). 

Our results for V836 Tau are intriguing in this context. 
Compared to the SEDs of well-studied transition objects 
such as GM Aur and DM Tau, where a strong infrared excess 
appears only beyond $\sim 10\micron$ indicating an optically 
thin region 3--20\,AU in size (Calvet et al.\ 2005), 
the SED of V836 Tau 
shows a significant infrared excess at a shorter wavelength 
$\sim 5-10\micron$. 
Therefore, if the V836 Tau system has an optically thin inner 
region, it is comparatively small and plausibly within the 
range of disk radii probed by CO 
fundamental emission (within $\lesssim 2$\,AU; Najita et al.\ 2003).  
The SED of V836 Tau can be fit with a 
simple model of a flared optically thick disk with an 
inner hole $\sim 1$\,AU in radius (see Appendix). 
In comparison, the SED of the transition object LkCa\,15 
can be interpreted as indicating an optically thin region 
within $\sim 3$\,AU (Bergin et al.\ 2004), 
or a radial gap extending from 5--46\,AU (Espaillat et al.\ 2007). 
A planet orbiting at such large distances ($\sim 3$\,AU or $\sim 40$\,AU)
may create a gap in the disk, but not 
radially truncate the gaseous disk significantly within 1\,AU.  
In contrast, systems like V836 Tau, in which the SED may indicate 
an optically thin region at much smaller radii ($< 1$\,AU), 
are the ones for which the CO fundamental emission would 
in principle be capable of diagnosing a physically truncated 
inner disk if an orbiting companion is present.
The radially truncated CO emission that we observe may 
support this picture. 

As an alternative interpretation of the available data, 
the short wavelength SED of V836 Tau 
($\lambda < 10\micron$) 
can also be fit with a simple model of an inclined ($i=60$), 
geometrically flat disk (see Appendix) that possibly 
results from significant grain growth and settling.
In such a situation, the gaseous disk would be radially continuous 
and we might expect to observe a CO profile like those of other 
non-transition T Tauri stars.  Such a profile, typically 
centrally peaked, is not observed.  

In the photoevaporation
and massive giant planet 
scenarios for the origin of a transitional SED, 
little gas is expected to be present within $\Rhole$, 
in contrast to the observed situation where a gaseous disk 
is present at 0.05--0.4\,AU and ongoing (possibly intermittent) 
stellar accretion is observed.  The relatively high 
disk mass of V836 Tau ($\sim 0.01\Msun$; Andrews \& Williams 2005), 
compared to the much smaller masses at which photoevaporation 
is expected to be able to create an inner hole 
($\sim 0.001\Msun$), further argues against the photoevaporation 
scenario.  The possibility of a massive planet is also 
restricted by current limits on the stellar radial velocity 
of V836 Tau, which constrain the mass of a companion within 
$0.4 - 1$\,AU to $<5-10 M_J$ (L. Prato, personal communication). 

These arguments are schematic in that they rely on theoretical
predictions that have not been verified observationally.  For
example, the sizes of gaps that will be induced by an orbiting
companion of a given mass, and the extent to which stellar accretion
will be reduced, are poorly known from an observational point of
view.  Stellar companions have been found to establish large inner
holes and to terminate stellar accretion in some transition objects
(CoKu Tau/4---Ireland \& Kraus 2008; D'Alessio et al.\ 2005) and
not in others (CS Cha---Guenther et al.\ 2007; Espaillat et al.\
2007).  The situation for lower mass companions is essentially
unexplored.  As theoretical predictions are tested, it will
be useful to examine in detail the range of companion masses that
are consistent with the properties of V836 Tau.

\section{Summary and Future Directions}

V836 Tau has been classified as a transition object (Strom et al.\ 1989), 
a system that may be on the verge of dissipating its disk, 
possibly as a consequence of planetesimal formation or giant 
planet formation.  These processes can reduce the continuum 
opacity in certain regions of the disk, producing an optically 
thin inner hole or a low column density gap. 
If V836 Tau has such an optically thin region, the weak near-infrared 
excess and the stronger $10\micron$ excess in the system indicates 
that the optically thin region is much smaller ($< 1$\,AU) than the 
optically thin region in well-studied transition objects such as 
GM Aur and DM Tau (Calvet et al.\ 2005), where the optically 
thin region is 3--20\,AU in radius.  
Thus CO fundamental emission, which probes the region $\lesssim 2$\,AU 
(Najita et al.\ 2003), can potentially map out the radial structure of 
the inner gaseous disk in V836 Tau to determine, for example, if  
the inner disk has been truncated by a companion orbiting within 1\,AU. 

Along these lines, we find that the $v$=1--0 CO fundamental line 
profiles of V836 Tau are unusual compared to those of 
other T Tauri stars in being markedly double-peaked. 
The strength and shape of the line emission is consistent with  
emission from a Keplerian disk over a limited range of 
radii ($\sim 0.05 - 0.4$\,AU). 
Further work is needed to determine whether the outer radius 
of the emission results from the physical truncation of the disk 
beyond $\sim 0.4$\,AU or the truncated excitation of the 
$v$=1--0 CO fundamental transitions beyond this radius.

A theoretical approach to this problem requires 
studies of the thermal, chemical, and excitation 
structure of disks at a level of detail that is appropriate for 
comparison with observations.  Studies of the possibility 
of non-LTE CO level populations in low accretion rate systems 
would be particularly welcome.  
For a more empirical approach to the problem, we might compare  
the observed line profiles of V836 Tau with other low accretion 
rate systems in which the disk is expected to be radially continuous 
within a few AU.  
As an example of the latter approach, we discussed the 
CO fundamental line profiles of LkCa15, a T Tauri star with a 
low accretion rate similar to that of V836 Tau.
The more centrally peaked line profiles of LkCa15, if 
representative of other low accretion rate systems, 
would suggest that the double-peaked emission profiles of 
V836 Tau arise from a physically truncated inner disk. 

Such a physically truncated inner disk might arise if the 
system has formed a Jovian mass planet that has cleared a gap in 
the disk.  A simple fit to the SED of V836 Tau is 
consistent with a flared disk that has an optically thin 
region within $\sim 1$\,AU.
Thus, a possible interpretation of the data is that 
an orbiting companion has created a gap between 
a gaseous inner disk within 0.4\,AU 
and an optically thick outer disk beyond 1\,AU. 
Since the above fit to the SED is non-unique, it would be 
useful to use both improved SED modeling techniques 
(e.g., Calvet et al.\ 2005) 
and infrared interferometry (e.g., Ratzka et al.\ 2007) 
to test the hypothesis that the dust disk has an 
optically thin gap or inner hole.  

In systems that have formed a Jovian mass planet, 
small grains may be filtered out of the inward accretion flow 
at the outer edge of the gap (Rice et al.\ 2006), rendering the 
dust distribution a poor tracer of the physical structure 
of the disk at smaller radii. 
Gaseous disk tracers, like the CO fundamental emission 
discussed here, may then be {\it needed} to probe disk structure 
at these smaller radii.
Thus, there is considerable motivation to expand the study of 
gaseous disk diagnostics beyond the present case, to 
understand more generally 
whether and how well diagnostics such as CO fundamental emission 
can probe the radial structure of gaseous disks.






\acknowledgments

We are grateful to Steve Strom for stimulating and 
insightful discussions on this topic.  
We also thank Lisa Prato for communicating her radial velocity 
results in advance of publication. 
Financial support for this work was provided by the NASA Origins 
of Solar Systems program (NNH07AG51I) and 
the NASA Astrobiology Institute
under Cooperative Agreement No.\ CAN-02-OSS-02 issued through the
Office of Space Science.  This work was also supported by the Life and
Planets Astrobiology Center (LAPLACE).
Basic research in infrared astronomy at the Naval Research Laboratory
is supported by 6.1 base funding. 
The authors wish to recognize and acknowledge the very significant
cultural role and reverence that the summit of Mauna Kea has always
had within the indigenous Hawaiian community.  We are most fortunate
to have the opportunity to conduct observations from this mountain.

\appendix

\section{Radial Dust Distribution}

The low stellar accretion rate of V836 Tau 
($\sim 10^{-9}\Msunperyr$; from Hartigan et al.\ [1995], 
scaled downward by a factor of $\sim 10$ as in Najita et al.\ [2007b] 
or Gullbring et al.\ [1998])
suggests that the disk temperature distribution will be 
dominated by the passive reprocessing of stellar radiation. 
Accordingly, simple SEDs fits suggest that the short 
wavelength SED can be fit either as a passive optically thick 
disk that is flat at small radii and radially continuous (Fig.~8) 
or as a passive, flared optically thick disk with an inner hole 
(Fig.~9).  

In the former case, the $3-8\micron$ IRAC excesses are well fit 
with a continuous flat disk, observed at an inclination $i=60$, 
that has a temperature distribution $T_D \propto r^{-3/4}$
with a normalization appropriate to a passive reprocessing disk 
(e.g., Adams, Lada, \& Shu 1988).
The disk is assumed to be optically thick beyond the 
dust sublimation radius (where $T_D = 1500$\,K) 
and to have negligible optical depth within that radius 
($<0.017$\,AU). 
While the $3-9\micron$ excess is well fit and arises from 
disk radii within 0.5\,AU, the excesses at 24 and 70$\micron$ 
are underpredicted.  A disk that is more strongly flared beyond 
$\sim 0.5$\,AU would likely produce a better fit. 
While such a simple model does a reasonable job fitting the 
short wavelength SED, it implies that the inner radius of 
the dust disk (at 0.017\,AU) is within the inner radius of the 
CO emission (at $\sim 0.05$\,AU; \S3.3).  This is in contrast 
to the situation found for more active T Tauri stars, where 
the SED is better fit with a frontally illuminated hot inner 
dust rim located further from the star (at $\gtrsim 0.1$\,AU). 
Such structures can account for the magnitude of the 
near-infrared excesses of T Tauri stars (Muzerolle et al.\ 2003).  
They also agree with the dust inner radii measured using 
infrared interferometry (e.g., Eisner et al.\ 2005).

To illustrate the latter case, we fit the SED using the 
public domain version of the CGPLUS modeling program 
written by C.\ P.\ Dullemond, C.\ Dominik, and A.\ Natta 
(http://www.mpia-hd.mpg.de/homes/dullemon/radtrans/) to calculate 
the disk contribution. 
The CGPLUS model is based on the models of Chiang \& Goldreich (1997) 
and Dullemond, Dominik, \& Natta (2001). 
The model includes four physical components: 
a stellar blackbody, a puffed-up inner rim, 
the disk surface, and the disk interior. 
Figure 9 shows the contribution to the SED of a disk truncated 
at an inner rim temperature of 330\,K, 
corresponding to a disk radius of 1.1\,AU. 
The inner rim (long dashed line) 
is modestly flared ($\chi_{\rm rim}=1.25$),  
and the rest of the disk is less strongly 
flared ($\chi_{\rm disk} = 0.6$).
While the stellar blackbody was used in calculating 
the temperature of the other components, in 
constructing the composite SED we used a 
Basel stellar atmosphere v2.2 (corrected; Lejeune 2002) 
of the same effective temperature. 
Adding a hot blackbody component $T=1500$\,K (short dashed line), 
representing hot dust located close to the star,  
produces a reasonable fit to the SED. 
There is significant degeneracy in the model parameters, as might be 
expected given the number of parameters in the CGPLUS 
model and the limited number of data points used in the fit.  
More restrictive fits could be obtained with a more sophisticated 
disk atmosphere model (e.g., D'Alessio et al.\ 2005),  
a self-consistent treatment of the inner and outer disk components,   
and constraints from interferometry (e.g., Ratzka et al.\ 2007).

\clearpage
\begin{figure}
\epsscale{0.9}
\plotfiddle{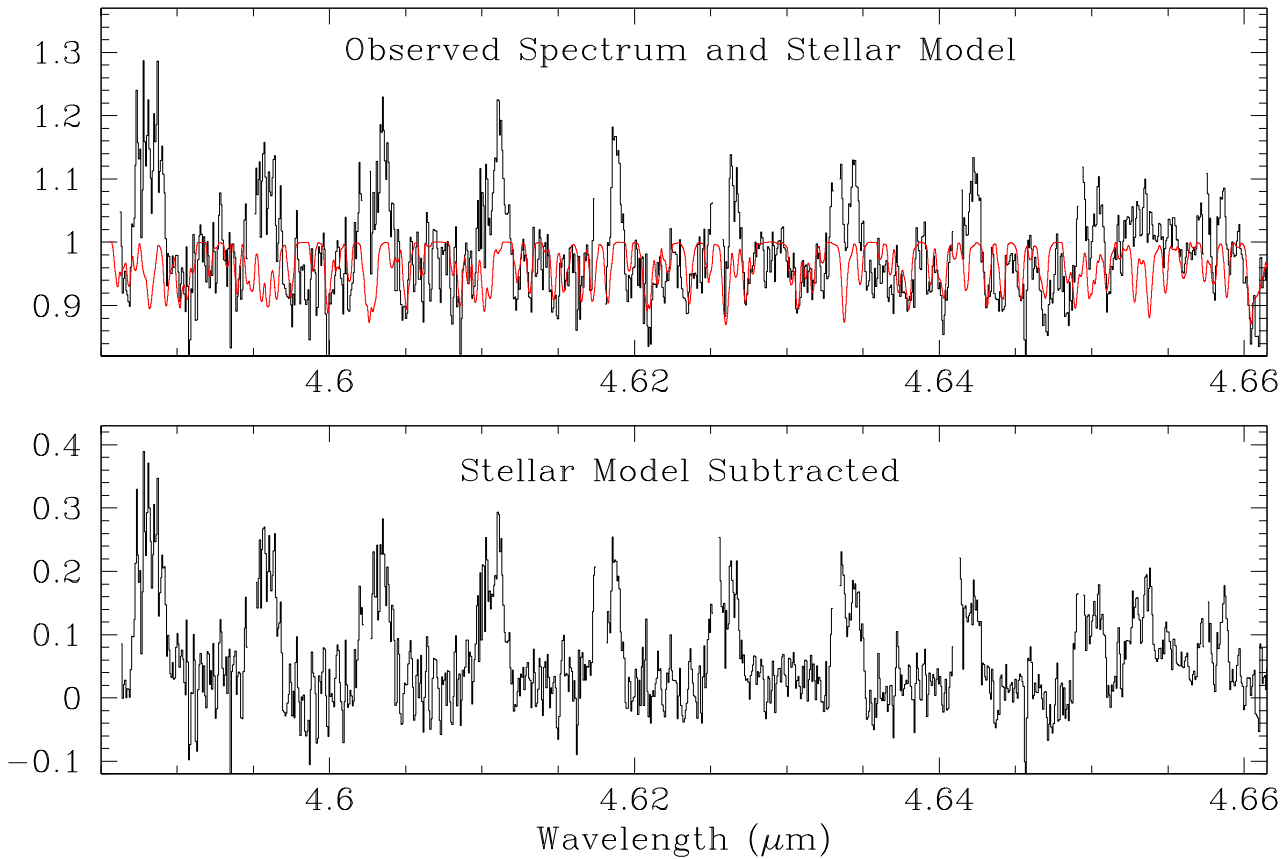}{5.5truein}{270}{75}{75}{-290}{430}
\caption{Observed spectrum of V836 Tau in the $4.6\micron$ region, normalized 
to the continuum level (top panel; histogram).  The adopted synthetic 
stellar photospheric spectrum with a veiling continuum level that is constant 
with wavelength is also shown (red line).  
The CO emission spectrum from the disk (in continuum flux units; bottom panel) 
is obtained by subtracting the veiled stellar photosphere  
from the observed spectrum.  
In both panels, regions of the spectrum that have poor telluric correction 
are not plotted.
Note that hydrogen 7--5 Pf$\beta$ absorption was present in the 
telluric standard used to calibrate the data.  Since we did not 
attempt to correct for the absorption, the hydrogen line emission 
at 4.654$\micron$  is artificially enhanced.
}
\end{figure}

\begin{figure}
\epsscale{0.9}
\plotfiddle{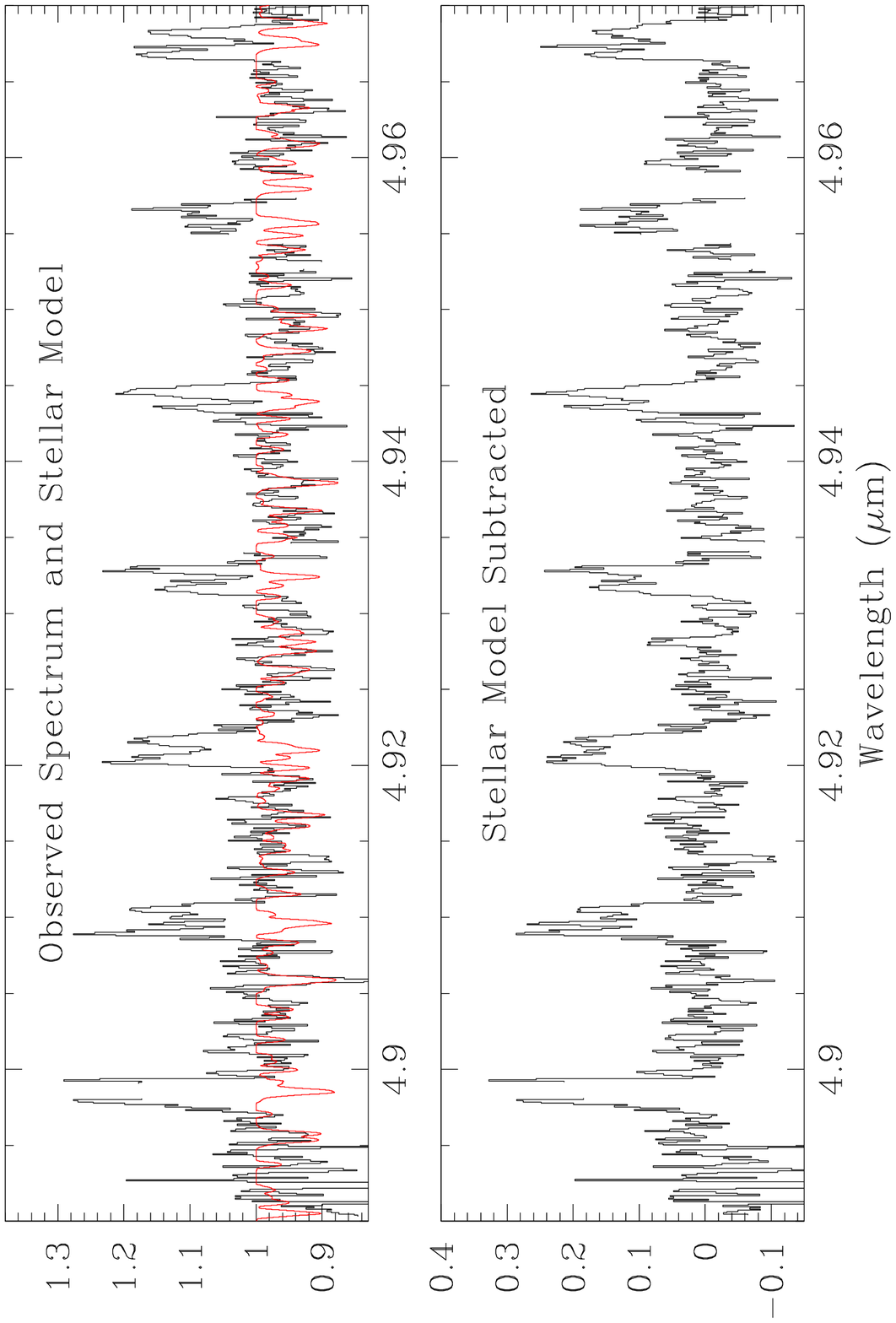}{4.5truein}{270}{75}{75}{-290}{430}
\caption{As in Figure 1, but for the $4.9\micron$ region. 
}
\end{figure}

\begin{figure}
\epsscale{0.9}
\plotone{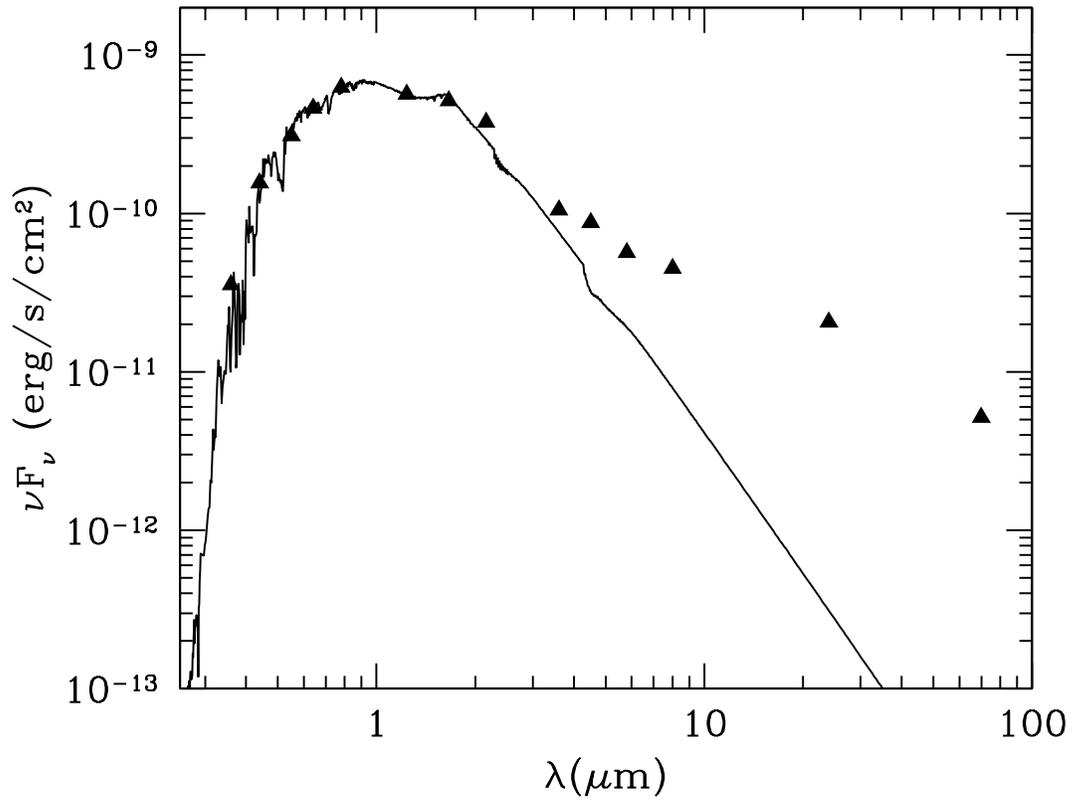}
\caption{SED of V836 Tau based on photometry in the literature 
and dereddened assuming $A_V=1.1$ (triangles; see text for 
details).  A Basel model atmosphere (v2.2 corrected) 
for $T_{\rm eff}=4060$\,K and $\log g=4.0$ (solid line) 
is shown for comparison. 
}
\end{figure}

\begin{figure}
\epsscale{0.9}
\plotone{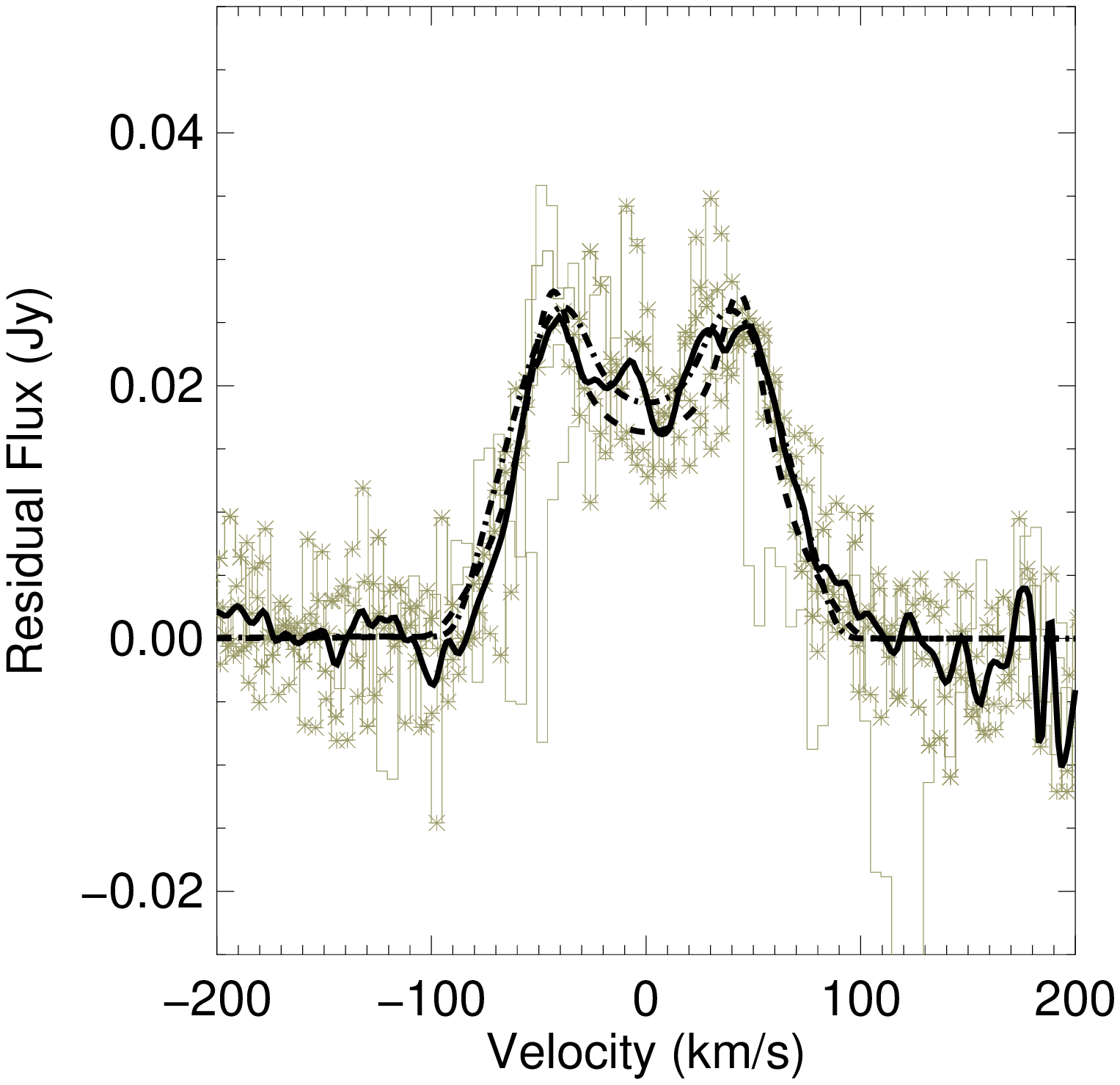}
\caption{Average CO emission profile from V836 Tau in the $4.9\micron$
region (heavy solid black line) compared with a synthetic disk emission profile
for the $v$=1--0 P26 line in a model that has an outer radius 
to the emission (dashed line; see Figure 6 and text for details).  
The synthetic P26 line profile for a model with a steep temperature 
gradient (dashed-dot line) is also shown (see text for details).
The velocities shown are relative to the V836 Tau stellar velocity. 
The individual lines that contribute to the average (1--0 P25 through P28 
and P30) are shown as gray histograms.  
Points included in the average are indicated by asterisks.  
The average profile is symmetric, double-peaked, and centered at the 
stellar velocity.
} 
\end{figure}

\begin{figure}
\epsscale{0.9}
\plotone{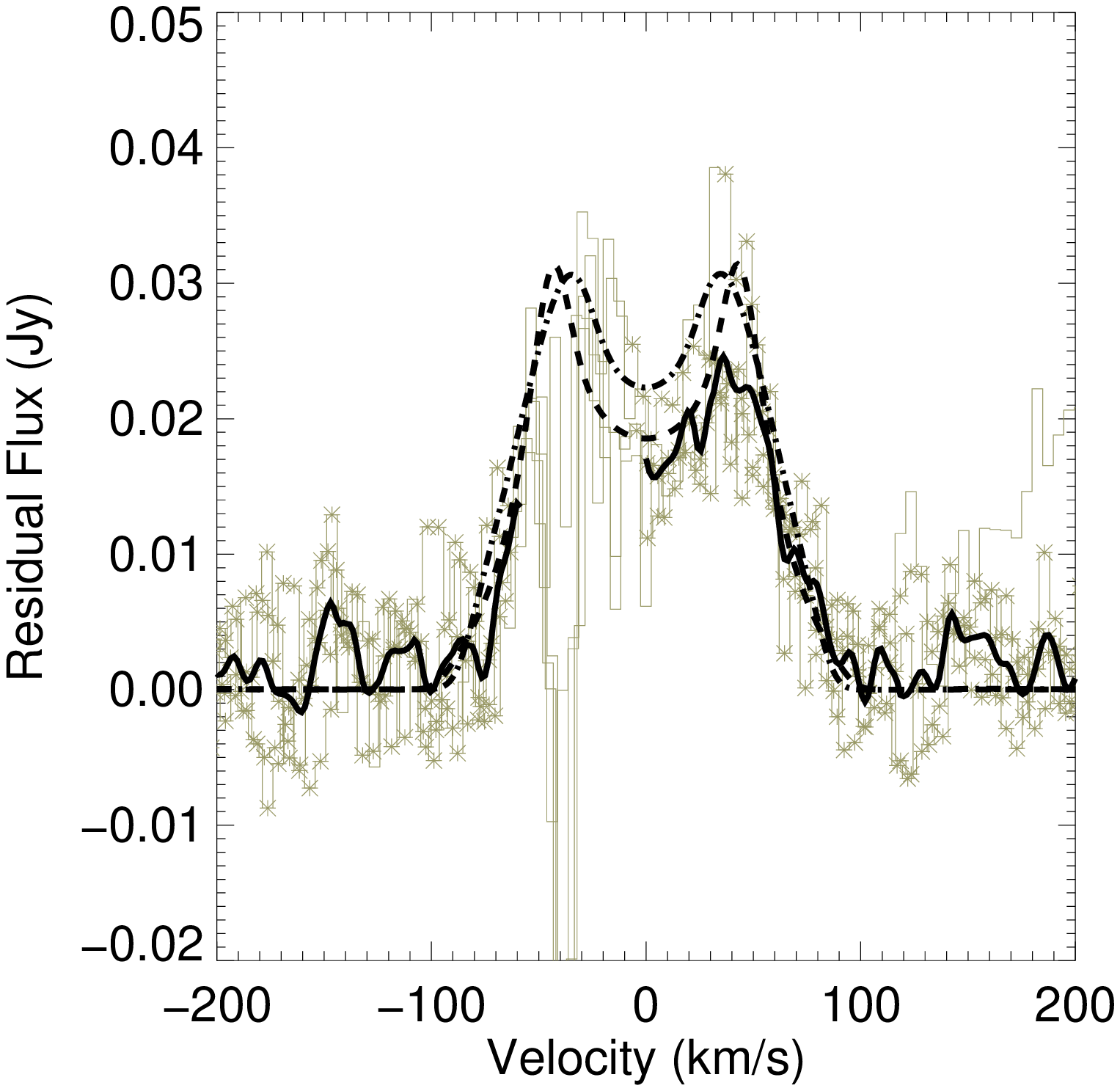}
\caption{Average CO emission profile from V836 Tau in the $4.6\micron$ 
region (heavy solid black line) compared with a synthetic disk emission profile
for the $v$=1--0 R3 line in a model that has an outer radius 
to the emission (dashed line; see Figure 6 and text for details).  
The synthetic R3 line profile for a model with a steep temperature 
gradient (dashed-dot line) is also shown (see text for details).
The velocities shown are relative to the stellar velocity of V836 Tau. 
The individual lines that contribute to the average 
(1--0 R1 through R4 and R6) are shown as gray histograms.  
Points included in the average are indicated by asterisks.  
Note that the region around the Pf\,$\beta$ line, located next to 
the 1--0 R1 line, was excluded from the average.  Strong telluric 
CO absorption at $-36\kms$ is present in all of the profiles, 
so there is no average line profile in the surrounding 
velocity interval.
} 
\end{figure}

\begin{figure}
\epsscale{0.9}
\plotfiddle{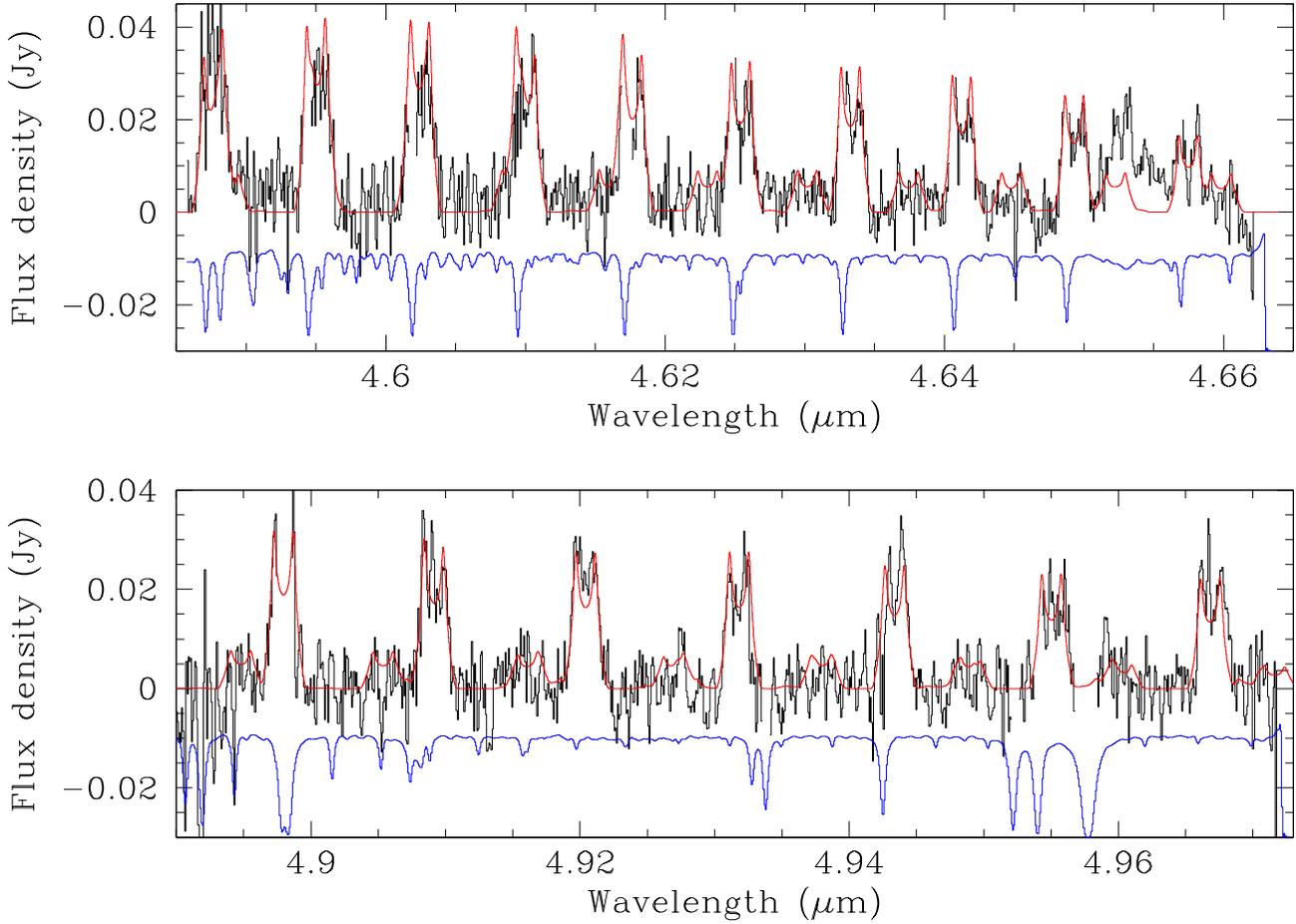}{5.0truein}{270}{75}{75}{-285}{430}
\caption{CO emission spectrum from V836 Tau (black histogram)
in the $4.6\micron$ (top) and $4.9\micron$ (bottom) regions 
compared with a model disk emission spectrum (red line) 
in which the emission is truncated at $\Rout=5\Rin$ 
(see text for details). 
Regions of poor telluric correction, as indicated by the telluric 
transmission spectrum (blue line), have been excised from the 
plotted observed spectrum. 
}
\end{figure}

\begin{figure}
\epsscale{0.9}
\plotfiddle{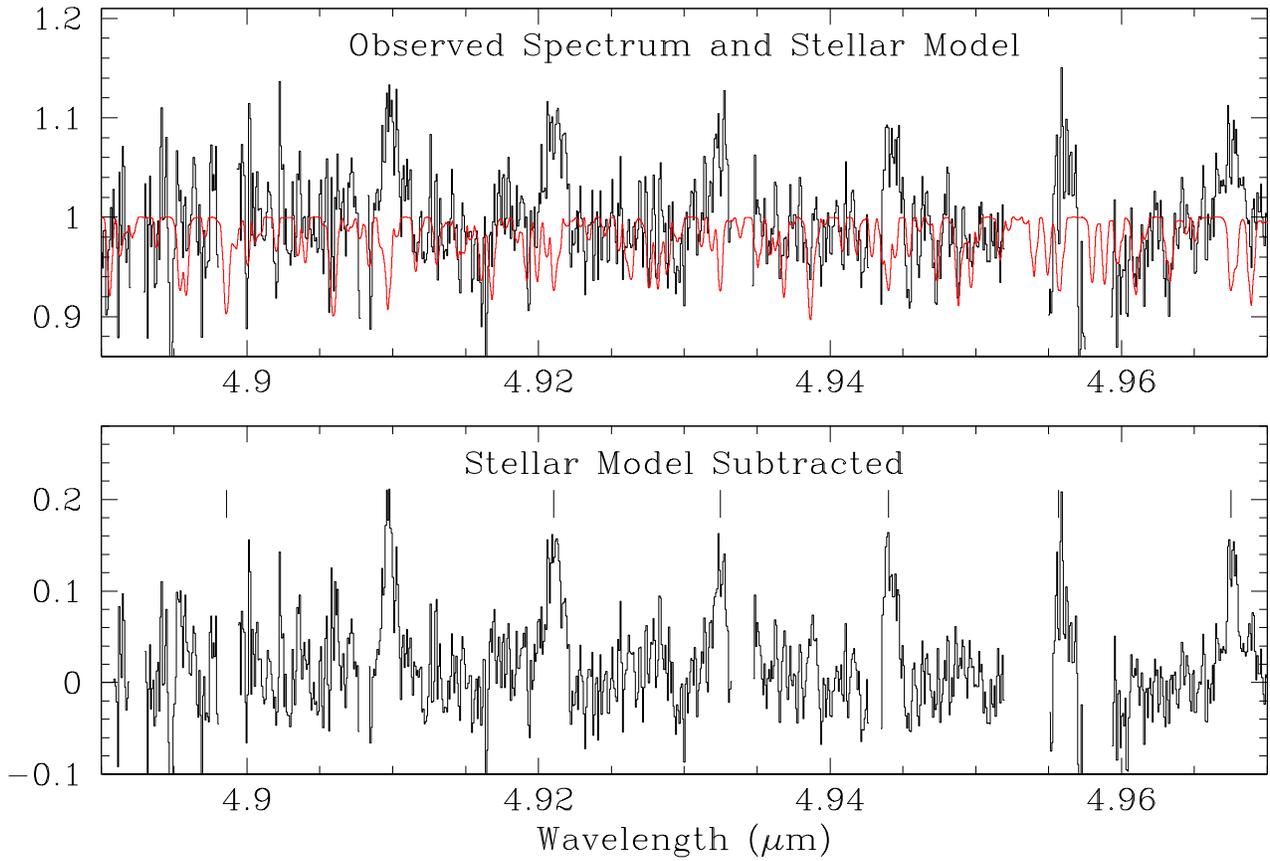}{5.5truein}{270}{75}{75}{-285}{430}
\caption{As in Figure 2, but for LkCa\,15. 
The wavelengths of CO $v$=1--0 lines at the 
stellar velocity of LkCa 15 are indicated in the bottom panel 
(short vertical lines). 
While LkCa\,15 is similar to V836 Tau in having a relatively 
low stellar accretion rate, its CO line profile is centrally 
peaked rather than double-peaked.
}
\end{figure}

\begin{figure}
\epsscale{0.9}
\plotone{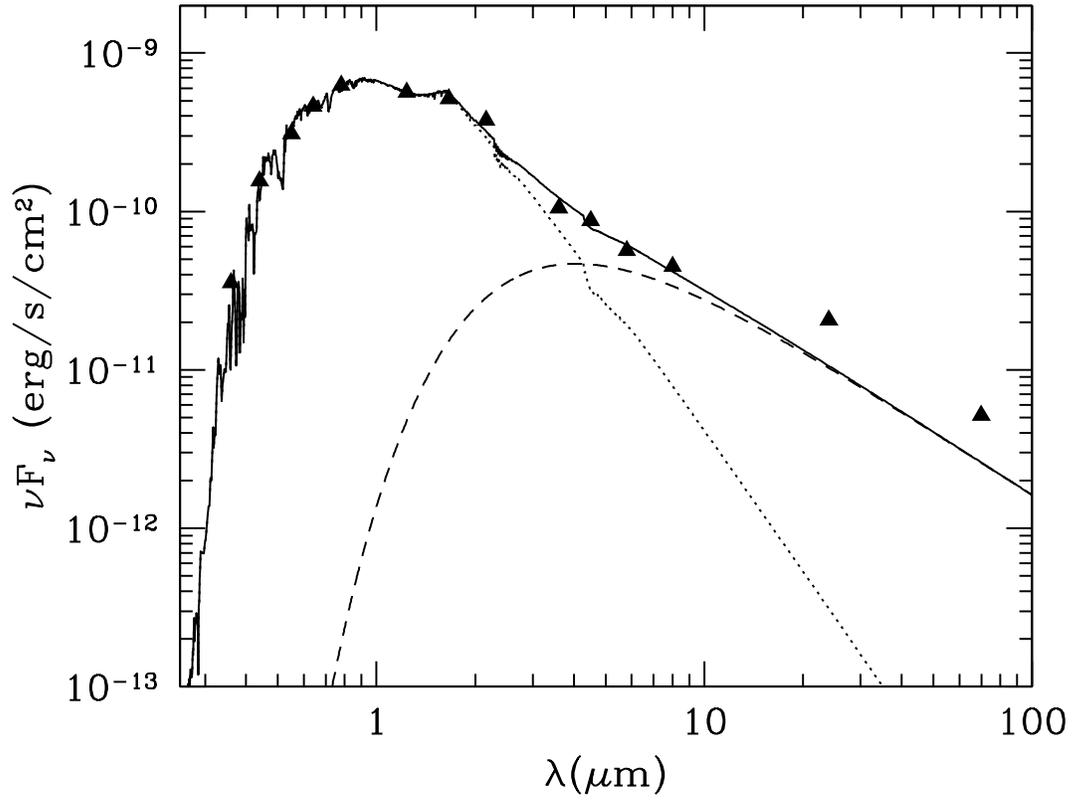}
\caption{A flat reprocessing disk (dashed line), 
combined with a model stellar photosphere (see Fig.~6; dotted line), 
and observed at $i=60$ 
provides a good fit to the dereddened SED of V836 Tau (triangles) 
at short wavelengths. 
}
\end{figure}

\begin{figure}
\epsscale{0.9}
\plotone{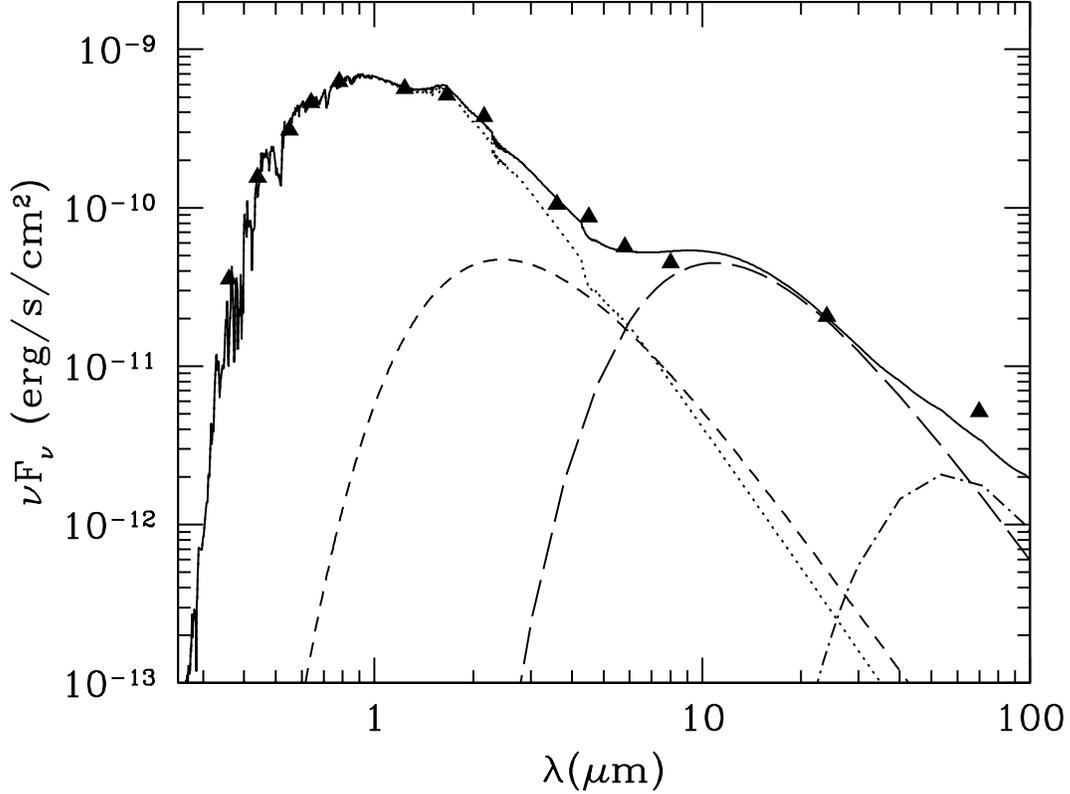}
\caption{
A reasonable fit to the dereddened SED of V836 Tau (triangles)  
is also obtained by combining 
a model stellar photosphere (see Fig.~6; dotted line), 
with the SED of a flared reprocessing 
CGPLUS disk that is truncated at 1.1\,AU,  
and a 1500\,K blackbody that represents residual dust within 1\,AU 
(short dashed line). 
An inclination of $i=60$ is assumed. 
The CGPLUS model includes contributions from the disk wall (long dashed line), 
the disk surface (dashed-dot line), and 
the disk midplane (not shown).  
The model assumes that the scale height of the disk is $\chi$ times the 
pressure scale height where $\chi = 1.25$ at the wall and $\chi=0.6$ 
at larger disk radii.
}
\end{figure}

\end{document}